\documentclass[12pt]{iopart}
\usepackage{graphicx}

\begin{document}

\title[From quantum feedback to probabilistic error correction]{From quantum
  feedback to probabilistic error correction: Manipulation of quantum
  beats\\ in cavity QED.}

\author{P Barberis-Blostein}

\address{Instituto de Investigaciones en Matem\'aticas Aplicadas y en
  Sistemas, Universidad Nacional Aut\'onoma de M\'exico, Ciudad Universitaria,
  04510, M\'exico, DF, M\'exico}

\author{D G Norris and L A Orozco}

\address{Joint Quantum Institute, Department of Physics, University of
  Maryland and National Institute of Standards and Technology, College
  Park, MD 20742, USA}

\author{ H J Carmichael }

\address{Department of Physics, University of Auckland, Private Bag
  92019, Auckland,\\ New Zealand}

\ead{lorozco@umd.edu}
\begin{abstract}
It is shown how to implement quantum feedback and probabilistic error
correction in an open quantum system consisting of a single atom, with ground-
and excited-state Zeeman structure, in a driven two-mode optical cavity. The
ground state superposition is manipulated and controlled through conditional
measurements and external fields, which shield the coherence and correct
quantum errors. Modeling of an experimentally realistic situation demonstrates
the robustness of the proposal for realization in the laboratory. 

\end{abstract}

\maketitle

\section{Introduction}
The conditional dynamics of quantum systems lie at the heart of quantum
feedback \cite{wisemanbook} and probabilistic error correction
\cite{koashi99}. Realizations of the latter in superconductor
\cite{korotkov06,katz08} and photonic \cite{kim09} qubits combine weak
measurements with reversal operations to recover an original quantum
state. The experiments of Smith {\it et al.} \cite{smith02,reiner04a}, which
employ strong quantum feedback, rely on knowledge of the conditional dynamics
of the system to capture and subsequently release a quantum state. Dissipation
and detection are critical to the implementation of the capture and release.

In this paper we study possible realizations of quantum feedback and
probabilistic error correction in an optical cavity QED system, an open
quantum system that is an important candidate for realistic implementations of
quantum information processing, in particular, the interconversion between
stored atomic qubits and ``flying'' qubits of light \cite{cirac97}.

The elements of the system are a single atom (or more generally atoms) which
has Zeeman structure in its ground and excited states. The atom interacts with
two orthogonal linear polarization modes of a high finesse cavity. One mode
excites $\pi$ transitions in the atom, and classical driving of this mode
introduces energy into the system. Some of this energy, through excitation and
decay of the atom, is transferred to the orthogonal mode.  The
  cavity, drive and $\pi$ transition frequencies are assumed to be the
  same.  With this
background, the conditional dynamics of interest work as follows: assuming an
atom prepared in the $m=0$ ground state, the detection of a photon in the
orthogonal mode sets the atom in a superposition of $m=\pm1$ ground states,
since it is not known whether the photon had $\sigma^{+}$ or $\sigma^{-}$
polarization; the prepared superposition then evolves in a magnetic field,
thus acquiring a relative phase, until another $\pi$ excitation transfers the
developed ground-state coherence to the excited state; subsequently, detecting
a second orthogonal-mode photon projects the atom into its starting state. The
sequence overall realizes the elements of a quantum eraser \cite{scully82}.

\section{System and its quantum beats}
\begin{figure}[h]
\begin{center}
\includegraphics*[width=4in]{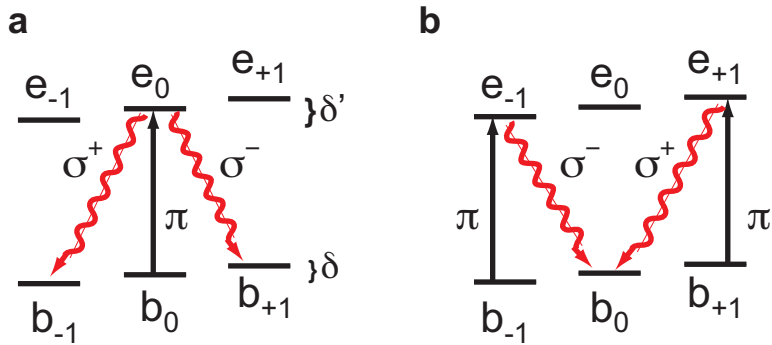}
\caption{\label{fig:atomo6niv} Model six-level atom, with $\pi$ transitions
  (straight lines) and $\sigma$ transitions (diagonal lines) displayed. The
  applied magnetic field  creates ground- and excited-state frequency
  detunings, $\delta$ and $\delta^\prime$, respectively, between the $m=\pm 1$
  and $m=0$ Zeeman sublevels. }
\end{center}
\end{figure}

We consider a two-mode driven optical cavity QED system in the regime of
intermediate coupling, where the dipole coupling constant is comparable to the
cavity and spontaneous emission decay rates.  In our model, we limit the
atomic basis to the six states of Figure~\ref{fig:atomo6niv}, for some
  transition $F\rightarrow\,F'$, where $F\neq F'$. This truncated basis
yields the simplest model to capture the physics of our proposal. We assume
that the ground- and excited-state Zeeman shifts are equal, {\it i.e.},
$\delta=\delta^\prime$, except where otherwise noted.  The magnetic field
  is perpendicular to the cavity axis, in the same direction as the
  polarization of the driven (V) mode.

Optical pumping prepares the atom in the state $|b_0\rangle$, after which, due
to its interaction with the driven cavity mode, it is excited to the state
$|e_0\rangle$. From here it may decay back to the ground state through the
emission of a $\pi, \sigma^{+}$, or $\sigma^{-}$ photon. If the emission is of
a $\pi$ photon, the emitted photon is added to the driven (V) mode; if,
however, it is of a $\sigma^{+}+\sigma^{-}$ photon, it will populate the
undriven (H) mode.  In the latter case, assuming broadband detectors, when the photon is detected (measured
in the H/V basis), the atom arrives in the superposition state
\begin{equation}
\label{eq:basesuperposition}
|\psi\rangle=(|b_{-1}\rangle+|b_1\rangle)/\sqrt{2}\, ,
\end{equation}
since neither the helicity nor the frequency of the photon is
known. The coefficients of the superposition are the same because
  the probabilty of going from $|e_0\rangle$ to $|b_{-1}\rangle$ or
  $|b_1\rangle$ emitting an undriven photon is equal.  This is because the ground-
  and excited-state Zeeman shifts are equal within the bandwidth of the cavity and cavity, drive and
  $\pi$ transition frequencies are assumed to be the same.  Note that
with the cavity decay rate comparable
to the atom-cavity coupling, it is probable that the photon leaks from
the cavity where it is detected before it can be reabsorbed. The atom
now undergoes Larmor precession in the applied magnetic field while it
continues to interact with the driven cavity mode. In this way the
prepared ground state superposition appears in the excited state as
the superposition
\begin{equation}
|\psi^\prime\rangle=(e^{-i\phi}|e_{-1}\rangle+e^{i\phi}|e_1\rangle)/\sqrt{2}\,
\label{eq:superphase}
\end{equation}
with the phase $\phi$ due to Larmor precession. From here the atom can return
to its original state, $|b_0\rangle$, by emitting a second photon into the
undriven (H) mode.

The conditional detection of both undriven-mode photons amounts to measuring
the second-order correlation function, $g^{(2)}(\tau)$, of the H-polarized
cavity output. Two indistinguishable paths yield ``start'' and ``stop''
photons for the measurement:
$|b_0\rangle\rightarrow|e_0\rangle\rightarrow|b_1\rangle\rightarrow|e_1\rangle\rightarrow|b_0\rangle$
and
$|b_0\rangle\rightarrow|e_0\rangle\rightarrow|b_{-1}\rangle\rightarrow|e_{-1}\rangle\rightarrow|b_0\rangle$. Since
the phase advance along each path is different ($\pm\phi$), and the difference
grows in time, interference between the paths yields oscillations, quantum
beats, in $g^{(2)}(\tau)$. It is important to note that the quantum beats are
not visible in the mean transmitted intensity of the undriven mode; a similar
observation appears in Refs.~\cite{javanainen92,hegerfeldt93,patnaik99} but for slightly
different atomic or optical situations. They arise only when one conditions
the ``stop'' photon detection on the preparation of the ground state
superposition signaled by the ``start'' photon detection.

With our restriction to six atomic levels, the Hamiltonian for the atom
interacting with the two orthogonally polarized cavity modes, written in a
rotating frame, is
\begin{equation}
H=H_0+H_I+H_{cp}\, ,
\label{eq:hamiltoniano}
\end{equation}
with
\begin{eqnarray}
H_0&=&\hbar\delta^\prime(\sigma_{e_{-1}e_{-1}}-\sigma_{e_1e_1})
+\hbar\delta(\sigma_{b_{-1}b_{-1}}-\sigma_{b_1b_1})\, ,\\ H_I&=&-\hbar
g\mkern-2mu \left[c_0\mkern3mu\sigma_{e_0b_0}a_1
  +c_0'\mkern3mu(\sigma_{e_1b_1}+\sigma_{e_{-1}b_{-1}})a_1\right.
  \nonumber\\ &&\left.c_1\mkern3mu(\sigma_{e_0b_1}+\sigma_{e_{0}b_{-1}})a_2
  +c_1'\mkern3mu(\sigma_{e_{-1}b_0}+\sigma_{e_{1}b_{0}})a_2\right]+{\rm
  c.c.}\, ,
\label{eq:hamiltoniani}\\
H_{cp}&=&i\hbar\mathcal E(a_1^\dagger-a_1) +i\hbar\xi_b(\hat a_1^\dagger
a_2-\hat a_2^\dagger a_1)\, ,
\end{eqnarray}
where $\sigma_{ij}=|i\rangle\langle j|$ and $\mathcal{E}$ is the external
drive of the V-polarized cavity mode. The term proportional to $\xi_b$ models
cavity birefringence (in such a way that the coupled field amplitudes have the
same phase). Operators $a_1$ and $a_2$ annihilate a photon in the driven and
undriven modes, respectively, $g$ is the dipole coupling constant between the
atom and the cavity field and the $c$s and $c'$s are atom-field coupling
  coefficients.

Including now spontaneous emission and cavity decay, the master
equation in the Lindblad form is
\begin{eqnarray}\label{eq:master}
\dot\rho(t)&=&(1/i\hbar)[H,\rho]+\kappa L[a_1]+\kappa
L[a_2]\nonumber\\ &&+\frac{\gamma}{2}
L[c_0\mkern3mu\sigma_{b_0e_0}+c_0'\mkern3mu(\sigma_{b_1e_1}+\sigma_{b_{-1}e_{-1}})]\nonumber\\ 
&&+\frac{\gamma}{2}L[\,c_1\mkern3mu\sigma_{b_0e_{-1}}+c_1'\mkern3mu\sigma_{b_1e_{0}}\,]\nonumber\\ 
&&+\frac{\gamma}{2}L[\,c_1'\mkern3mu\sigma_{b_{-1}e_{0}}+c_1\mkern3mu\sigma_{b_{0}e_{1}}\,]\, ,
\end{eqnarray}
with
\[
L[o]=2o\rho o^\dagger-o^\dagger o\rho-\rho o^\dagger o\, ,
\]
where $\gamma$ is the spontaneous decay rate to modes other than the cavity
modes and $\kappa$ is the cavity field decay rate.

We
solve Eq.~(\ref{eq:master}) using a truncated photon basis and use the
so-called quantum regression theorem, in order to illustrate the main properties of the correlation function. We should note first that in
equilibrium---{\it i.e.}, considering the steady-state solution to
Eq.~(\ref{eq:master})---the populations in states $|b_0\rangle$ and
$|b_i\rangle$, $i=-1,1$, are all of the same order. In the limit of weak
driving, and assuming the atom moves very slowly across the cavity, the system
has three distinct, and widely different, time scales. The first, which
governs the long-time evolution of the master equation to its steady state, is
the time taken for the atomic ground states to reach equilibrium through
optical pumping; this time goes to infinity as the drive strength, $\mathcal
E$, approaches zero.  The second is the time for the cavity field to reach a
quasi-steady state, in equilibrium with the drive and the still slowly
evolving atomic state. This shortest of the three time scales is of order
$\kappa^{-1}$. The third, the intermediate time scale, is the time taken for
the atom to transit the cavity. Our calculations assume that the first photon
is detected after the transient on the shortest time scale has decayed---{\it
  i.e.}, after the cavity field reaches its quasi-steady state. Then, while
the atom transits the cavity, on the intermediate time scale, it is possible
to guarantee that a first H-polarized photon detection leaves the atom in the
superposition state of $|\psi\rangle$ of Eq.~(\ref{eq:basesuperposition}).

\begin{figure}[h]
\begin{center}
\includegraphics*[width=4in]{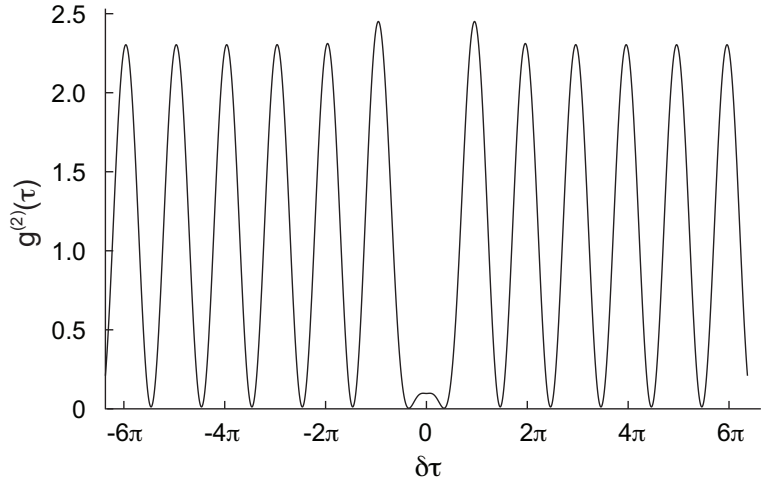}
\caption{\label{fig:fringe} Normalized second-order correlation function of
  the undriven mode. The atom interacts with a fixed dipole coupling constant,
  $g$, for $\kappa/\gamma=1/2$, $g/\gamma=1/4$, $\delta/\gamma=1/2$,
  $\mathcal{E}/\gamma=1/64$,
  $c_0=\sqrt{4/7}$,$c_0'=\sqrt{15/28}$,$c_1=-\sqrt{3/14}$,$c_1'=-\sqrt{5/14}$.
}
\end{center}
\end{figure}

Figure~\ref{fig:fringe} displays an example of the computed second-order
correlation function assuming a constant coupling between the atom and the
cavity modes; the correlation function decays to unity on the (longest)
optical pumping time scale.
The quantum beats due to Larmor precession are prominently displayed,
though they are suppressed around $\tau=0$ due to the antibunching of
single-atom fluorescence (it is impossible for one atom to emit two photons
within a single lifetime).  Note that on applying the interaction Hamiltonian,
Eq.~(\ref{eq:hamiltoniani}), to the state $|\psi^\prime\rangle$,
Eq.~(\ref{eq:superphase}), the coupling between the atom and the undriven
cavity mode vanishes for $\phi=(2n+1)\pi/2$, $n$ an integer; thus, an
H-polarized photon cannot be emitted at these times, the origin of the
interference zeros in Figure~\ref{fig:fringe}. For a Zeeman frequency
shift $\delta$, the phase gained through Larmor precession in time $\tau$ is
$\phi=\delta\tau$; hence the zeros occur for $\delta\tau=(2n+1)\pi/2$ with a
frequency $2\delta$. In Fig.~\ref{fig:fringe} the fringe visibility is
maximal, as the probabilities for finding the atom in state $|b_{-1}\rangle$
or $|b_1\rangle$ are the same within our model. Engineering a different
superposition, e.g.,
$|\psi\rangle=\alpha_{-1}|b_{-1}\rangle+\alpha_1|b_{1}\rangle$, with
$|\alpha_{-1}|\neq|\alpha_1|$, would yield a lower visibility.

Our proposal differs from work done with similar schemes \cite{wilk07,wilk07b,weber09}. There
  are differences in the configuration: we use a CW laser to pump the
  cavity, which then drives the atom, and a magnetic field perpendicular to
  the cavity axis so that $\sigma^+$ and $\sigma^-$ photons cannot be
  distinguished. There are also qualitative differences. Rather than focusing
  on the generation of entangled photons and the transfer of information
  between the atom and light, we focus on atomic state manipulation. For this
  reason it is crucial to disentangle the atom from the field. The field is
  used to prepare and read the atomic state. We prepare the state with a
  measurement: the first photon detection prepares the system in the 
  superposition, Eq. \ref{eq:basesuperposition}. This initialization disentangles the atom from the
  field and allows us, via feedback, to manipulate the atomic state alone. It
  shields the atomic state from alteration by a field measurement, making its
  manipulation less susceptible to errors. Instead of using quantum state
  tomography to infer the atomic state, we infer it from the phase and
  visibility of the second-photon fringe.

\section{Manipulation of the quantum beats}

\subsection{Feedback protocol}

Detection of the first photon prepares the superposition state which then
evolves in the magnetic field. Quantum feedback similar to that of
Refs.~\cite{smith02,reiner04a} requires one to stop the Larmor oscillation of
the atom in the ground state, and after a predetermined time continue it.

The quantum trajectory formalism \cite{carmichael93book} allows one to study
the dynamics of a dissipative quantum system in real time. We show in this
section what happens if, after a pre-set time, another laser beam excites one
of the two parts of the superposition to a different state. The excitation
interrupts the coherent evolution of the superposition. We envision it
executed by a pulse of light which takes the specific ground-state magnetic
sublevel to a different ground-state hyperfine (F) sublevel. The exciting
pulse arrives from a direction perpendicular to the cavity axis and lasts for
the appropriate time. The small detuning given by the magnetic field to the
ground-state sublevels is large enough for selective excitation with a narrow
laser or a two photon Raman transition. During the state transfer the external drive to the cavity is turned
off.

\begin{figure}[h]
\begin{center}
\includegraphics[width=4in]{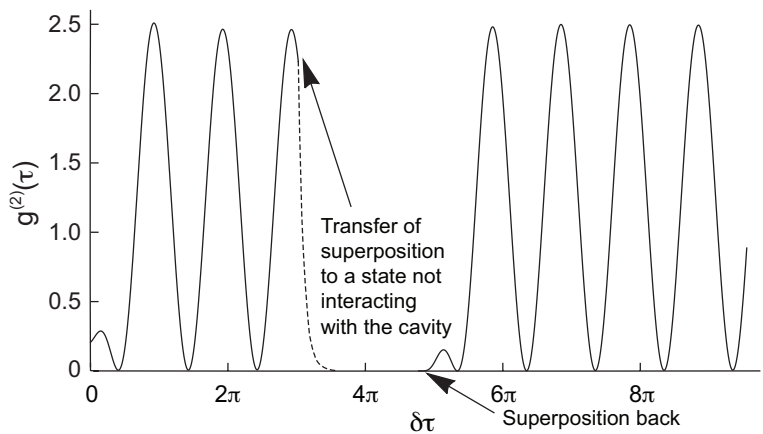}
\caption{\label{fig:feedback} Manipulation of the quantum beats with quantum
  feedback. The first arrow indicates the time when the atomic state
is transferred to a state that does not interact with the cavity modes. The
dotted line indicates the decay of the now undriven cavity field. The second
arrow indicates the time when the atomic state is brought back, and the
quantum beats resume. Parameters as in the previous figure.}
\end{center}
\end{figure}

There are two possible approaches to this excitation. One would be to excite
to a state that cannot return in any way to the superposition. This simply
interrupts the quantum beats so that the autocorrelation of the undriven mode
reveals a sudden change in the dynamics---the oscillations cease; the
interruption may be made at any time one chooses.  A more interesting
possibility is to use two Raman pulses to move the superposition in its entirety to another ground state while preserving the relative phase and with no spontaneous emission.  A second pair of Raman pulses brings it back to resume the quantum beats. Figure~\ref{fig:feedback} shows
the computed correlation function, where the first arrow marks the time of the
first state transfer and the second the time when the state is transferred back. 

\subsection{Probabilistic quantum error correction}

The previous subsection illustrated the possibilities for quantum feedback on
our system: one can interrupt the quantum beats and restart them later. We now
go further and look at the possibility of applying the basic protocol to
implement probabilistic error correction in Cavity QED. The idea developed
derives from recent experiments with superconducting qubits
\cite{korotkov06,katz08}. We first briefly describe how to correct an
  unwanted weak measurement using the probabilistic quantum error protocol
  described in Ref. \cite{koashi99} (see Ref.~\cite{kim09} for an optics
  application). Then we describe its implementation in our Cavity QED system.

Consider a qubit in the superposition
$|\psi\rangle=\alpha_0|0\rangle+\alpha_1|1\rangle$, corresponding to the state
of the atom after detecting a first undriven photon. If a weak measurement is
performed to detect state $|1\rangle$ (weak in the sense that if the qubit is
in $|1\rangle$, the probability $p$ to find the state is much less than
unity), the effect of not detecting $|1\rangle$ is to partially collapse the
superposition towards $|0\rangle$. In terms of the generalized measurement
operators \cite{nielsenbook}
\begin{eqnarray}
M_a&=&\sqrt{p}\mkern3mu|1\rangle\langle 1|\,
,\\ M_b&=&1-\left(1-\sqrt{1-p}\mkern3mu\right)|1\rangle\langle 1|\, ,
\end{eqnarray}
the probability of detecting state $|1\rangle$ is $p_1=\langle\psi|M_a^\dagger
M_a|\psi\rangle=p|\alpha_1|^2$, while the probability of not detecting it is
$p_0=\langle\psi|M_b^\dagger M_b|\psi\rangle=1-p|\alpha_1|^2$. After the
measurement, if we detect state $|1\rangle$, $|\psi\rangle$ collapses to
$|1\rangle$.  If we do not, we cannot deduce that the state of the system is
$|0\rangle$; nevertheless, some information has been gained. This is easily
seen from the fact that if we measure in a similar way an infinite number of
times and never find the state $|1\rangle$, the deduction, then, most surely,
is that the state is $|0\rangle$.  Thus, after a first measurement, if we fail
to find state $|1\rangle$, $|\psi\rangle$ is partially collapsed to
\begin{equation}
|\psi'\rangle=\frac{M_b|\psi\rangle}{\sqrt{\langle\psi|M_b^\dagger
    M_b|\psi\rangle}}=\frac{\alpha_0|0\rangle+\alpha_1\sqrt{1-p}|1\rangle}{\sqrt{1-p|\alpha_1|^2}}\, .
\end{equation}

The aim now is to correct the error introduced by the partial collapse. This
may be done by performing the following operations: we swap $|0\rangle$ and
$|1\rangle$ in $|\psi^\prime\rangle$ and repeat the measurement to detect
state $|1\rangle$. The probability of a null result is now
$(1-p)/(1-p|\alpha_1|^2)$, and again, given the null result the state
partially collapses to $|0\rangle$. Then swapping $|0\rangle$ and $|1\rangle$
for a second time the original state, $|\psi\rangle$, is recovered.  The
probability of recovery after the two measurements is $(1-p)$.

What if we do not know the outcome of the measurements? In this case the
density operator formalism can account for our ignorance. The total Hilbert
space is $H_q\otimes H_d$, where $H_q$ is the two dimensional Hilbert space of
the qubit and $H_d$ is the detector Hilbert space. The initial detector state
is $|{\rm no}\rangle$, and if it detects the state $|1\rangle$ it changes to
$|{\rm yes}\rangle$. The initial state of the qubit plus detector is
$|\psi\rangle|{\rm no}\rangle$. After the first measurement the state of the
system is
\begin{equation}
\rho=(1-p|\alpha_1|^2)\left(|\psi^\prime\rangle\langle\psi^\prime|\right)\left(|{\rm
  no}\rangle\langle {\rm
  no}|\right)+p|\alpha_1|^2\left(|1\rangle\langle1|\right)\left(|{\rm
  yes}\rangle\langle{\rm yes}|\right)\, ,
\end{equation}
and after the second measurement,
\begin{equation}
\rho'=(1-p)\left(|\psi\rangle\langle\psi|\right)\left(|{\rm
  no}\rangle\langle{\rm no}|\right)+p\left(|1\rangle\langle
1|\right)\left(|{\rm yes}\rangle\langle{\rm yes}|\right)\, .
\end{equation}
In this case the swapping of $|0\rangle$ and $|1\rangle$ must be carried out
independently of the measurement outcomes which are assumed unknown. Still,
there is a probability $1-p$ that the detector recorded two negative answers
and the system is in state $|\psi\rangle|{\rm no}\rangle$ and a probability
$p$ that it recorded a positive answer (at least one) and the system is in
state $|1\rangle|{\rm yes}\rangle$.

\subsection{Implementation in cavity QED}

In order to implement the protocol in cavity QED we define $|b_1\rangle$ and
$|b_{-1}\rangle$ as the states of the qubit. The measurement of a first
undriven mode photon then starts a series of pulses (e.g., inducing Raman
transitions between the qubit states) that initializes the state
\begin{equation}
|\psi\rangle=\alpha_0|b_{-1}\rangle+\alpha_1|b_1\rangle\, .
\end{equation}
In order to verify the preparation, the qubit state information may be mapped
in $H$-polarized photons, with the state of the qubit obtained from the
visibility and phase of the autocorrelation function. Using the predictability
$\mathcal{P}={\rm mod}(|\langle b_{-1}|\psi\rangle|^2-|\langle
b_{1}|\psi\rangle|^2)$ and the visibility $\mathcal{V}$, we have numerically
checked for the parameters used in this paper that
\begin{equation}\label{eq:predictability}
\mathcal{P}^2+\mathcal{V}^2\approx 1\, .
\end{equation}
This is an analog of the inequality characterizing a two-way interferometer
\cite{englert96}, and it allows us to determine the relative probability to
occupy the qubit states from the fringe visibility. The relative phase of the
state amplitudes is carried by the phase of the quantum beats. The weak
measurement might be implemented by a pulse of light, such that if the atom is
in state $|b_{1}\rangle$ there is a probability $p\ll1$ of it being
ionized. If the atom is ionized (probability $p |\alpha_1|^2$), the unitary
inverse operation is impossible, {\it i.e.}, the pulse makes a measurement
with outcome ``yes''. If we do not know the outcome, the system density
operator is
\begin{equation}
\rho=(1-p|\alpha_1|^2)\left(|\psi^\prime\rangle\langle\psi^\prime|\right)
\left(|{\rm no}\rangle\langle{\rm no}|\right)+p|\alpha_1|^2\left(|{\rm
  ion}\rangle \langle{\rm ion}|\right)\left(|{\rm yes}\rangle\langle{\rm
  yes}|\right)\, ,
\end{equation}
where $|{\rm ion}\rangle$ is the state of the ionized atom. If the atom is
ionized, the cavity no longer interacts with the atom and no second photon is
emitted. Otherwise, a second photon may be emitted, with the resulting quantum
beats having a visibility determined by the state $|\psi^\prime\rangle$.

The state $|\psi\rangle$ is recovered in the manner outlined above: after the
first weak measurement, we swap states $|b_1\rangle$ and $|b_{-1}\rangle$, and
then repeat the weak measurement and swap the states again. The final density
operator is
\begin{equation}
\rho^\prime=(1-p)\left(|\psi\rangle\langle \psi|\right)\left(|{\rm no}\rangle
\langle{\rm no}|\right)+p\left(|{\rm ion}\rangle\langle {\rm ion}|\right)
\left(|{\rm yes}\rangle\langle{\rm yes}|\right)\, .
\end{equation}
Note that with a probability $1-p$ the atom was not ionized and the state of
the system is $|\psi\rangle$.  If a second photon is detected, the
$g^{(2)}(\tau)$ constructed from these events will correspond to the original
state $|\psi\rangle$.  The visibility will be as if no weak measurements were
performed. The modification of the state introduced by the first measurement
is corrected by the second.  The probability for the scheme to be successful
is $1-p$. The visibility diminishes in the case where $|\psi\rangle$ has equal
coefficients if the protocol is not implemented after the first weak
measurement. The recovery of the initial maximal visibility is a necessary
signature that the protocol is working.

The experimental realization of the probabilistic quantum error correction
protocol in superconducting phase qubits~\cite{katz08} requires knowing the
probability of success of the partial measurement in order to correctly
estimate the recovered quantum state. In the presented scheme this is not
needed.

\section{ Experimental proposal and sensitivity analysis for atoms from an atomic beam}
\label{sec:sensitivity}
This section briefly describes the experimental apparatus where this proposal can be implemented, and analyzes the sensitivity of the measured correlation function to
certain experimental realities. Since the correlation function carries all the
information required for the proposed quantum feedback and error correction,
it is important to compute it for the realistic situation where atoms are
provided by an atomic beam. Such atoms are produced with random arrival times,
velocities, and directions of motion through the cavity mode function. We also
comment on a number of other experimental considerations that might diminish
the visibility of the quantum beats, or alter their frequency.

Our
experimental apparatus consists of a $2\mkern2mu{\rm mm}$ Fabry-Perot cavity
and a source of cold $^{85}$Rb atoms \cite{terraciano07a,terraciano09}. The
source delivers, on average, less than the equivalent of one maximally coupled atom
within the mode volume of the cavity at all times. The cavity supports two
degenerate modes of orthogonal linear polarization (H and V). During their
transit, the atoms interact with the orthogonally polarized modes for
approximately $5\mkern2mu\mu{\rm s}$. (See Refs.~\cite{birnbaum05,aoki09} for
similar systems.) In the presence of a weak magnetic field, and with the appropriate choice of
hyperfine levels, the atomic structure of Rb allows for the separation of
spontaneous emission events in the cavity. Driving $\pi$ transitions with V
polarization means that any light emitted along the cavity axis with H
polarization must come from spontaneous decay via a linear superposition of
$\sigma^{+}$ and $\sigma^{-}$ light.

We use in the atom-field interaction the Clebsch-Gordon coefficients coupling
the central six sublevels of the $D_2$ line of $^{85}$Rb on the $F=3$ to $F=4$
transition, {\it i.e.}, we impose a truncation of the full atomic raising and
lowering operators. The justification for this truncation is as follows. Atoms
interact with the cavity field for a finite time in the experiment. Their
coupling to the field changes as a function of position as they transit the
cavity. This limits the time over which a given atom can emit two photons,
imposing a limit on the duration of the observed quantum beats. Due to the
weak pump field, the period of a Rabi oscillation between the ground and
excited levels is much larger than the time the atom interacts with the
cavity. The spreading of the population to states outside the considered
manifold, before and after the first photon measurement, is negligible;
moreover, the proposed manipulations of the atomic state remains inside the
considered manifold.

In the weak limit at most two undriven photons are emitted by each atom inside the cavity. When the second photon is emitted from $\psi^\prime$, Eq.~(\ref{eq:superphase}), the atom can finish in a state different to $b_0$, thus reducing the fringe visibility. An estimation of how much the visibility decreases can be obtained from the Clebsch-Gordon coefficient that couples states $e_{\pm 1}$ with states $b_{\pm 2}$. The transition probability from $\psi^\prime$ to any of the ground states yields a visibility of $5/8$ instead of $1$. The fact that the fringe does not disappear altogether can be understood in the following way. When the phase of $\psi^\prime$ is $\phi=n\pi$, $n$ an integer---{\it i.e.} when the probability for a second photon emission is maximum---the photon leaves the truncated six-fold manifold with probability $3/13$; nevertheless, in the majority of the cases (probability $10/13$) it contributes to the quantum beats.

We model the source of atoms as a dilute atomic beam, assuming that
the cavity is either empty of atoms or occupied by one atom at the most.  The
atomic trajectories and interaction volume are defined in an manner similar to
that of Ref.~\cite{horvath07}. The mean intracavity photon number of the
undriven mode in steady state is
\begin{equation}
\overline{\langle a_2^\dagger
  a_2\rangle}=\overline{(\overline{N}/t_j)\int_0^{t_j} \langle a_2^\dagger
  a_2(t)\rangle d t}\, ,
\end{equation}
where $\overline{N}$ is the mean number of atoms in the interacting volume
(probability to find an atom in the cavity) and the integral performs a time
average over the passage of atom $j$ through the cavity mode function, with
transit time $t_j$ and time-dependent coupling $g_j(t)$. The overbar denotes
the average over an ensemble of atoms $\{j\}$. Similarly, we calculate the
unnormalized correlation function in steady state as
\begin{equation}
\overline{G^{(2)}(\tau)}=\overline{(\overline{N}/t_j)\int_0^{t_j} \langle
  a_2^\dagger(t)a_2^\dagger(t+\tau)a_2(t+\tau)a_2(t)\rangle dt}\, .
\label{eq:g2montecarlo}
\end{equation}

\begin{figure}[h]
\begin{center}
  \includegraphics*[width=5in]{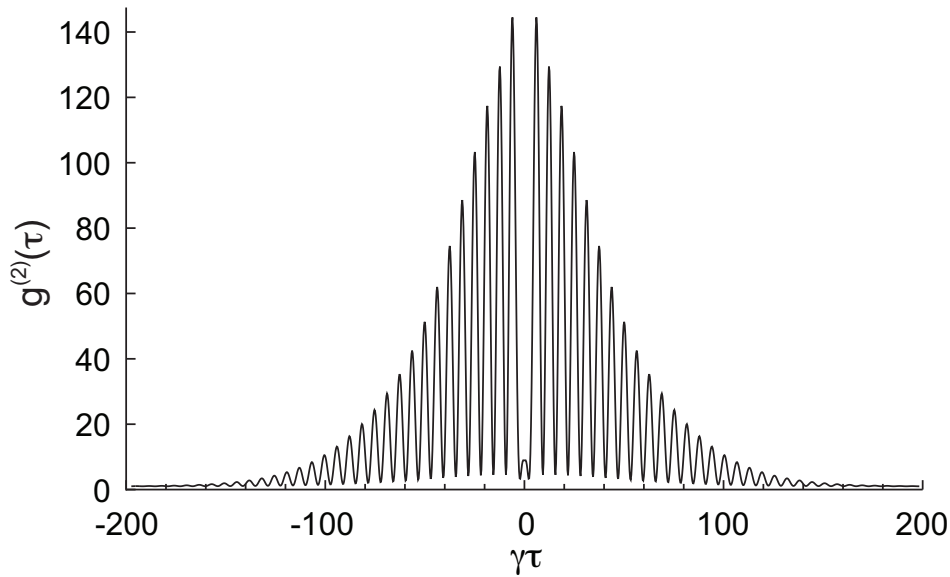}
  \caption{\label{fig:montecarlo} Montecarlo simulation of the undriven
    normalized second-order correlation function for a cavity with waist
    $w=56\mu$m and linewidth $\lambda=780$ nm. The Low-Velocity source of
    atoms has mean velocity $\overline{v}=15$ m/s with deviation $\Delta
    v=1.1$ m/s. The parallel and transverse (to the cavity axis) divergence
    angles are $\Delta\theta_p=1/40$ and $\Delta\theta_t=1/80$
    respectively. Other parameters: $\gamma/2\pi=6$ MHz, $\kappa/\gamma=1/2$, $g/\gamma=1/4$,
    $\delta/\gamma=1/2$, $\overline{N}=0.1$, $\mathcal{E}/\gamma=1/16$,
    $c_0=\sqrt{4/7}$, $c_0'=\sqrt{15/28}$,
    $c_1=-\sqrt{3/14}$, $c_1'=-\sqrt{5/14}$.}
\end{center}
\end{figure}

The transit time imposes a limit on the duration
of the observed quantum beats compared with the ongoing beats of
Fig.~\ref{fig:fringe}. So long as the external drive is weak, the atom transit
time is the only source of ``decoherence'' in the system; the finite lifetime
of the atomic excited state and of photons in the cavity does not limit the
duration of the quantum beats, nor does any optical pumping under the weak drive limit.

For each trajectory, the atom-cavity coupling---hence $G^{(2)}(\tau)$---goes
to zero as $\tau$ goes to infinity. In reality atoms enter and leave the
cavity continuously, so different atoms provide ``start'' and ``stop'' photons
at long delays and the correlation function does not go to zero. Assuming at
most one atom in the cavity at a time, and a mean time between atoms much
larger than the photon lifetime in the cavity, the normalized autocorrelation
function can be written as
\begin{equation}\label{eq:montecarloeq}
g^{(2)}(\tau)=1+\overline{G^{(2)}(\tau)}/\left(\overline{\langle a_2^\dagger
  a_2\rangle}\mkern3mu\right)^2\, ,
\end{equation}
where the background of unity comes from the correlation of photons emitted by
different atoms, while the term rising above unity correlates photons from the
same atom. Since the restriction to one atom at a time requires
$\overline{N}\ll1$, the so-normalized quantum beat has an amplitude much
greater than unity. 

Equation~\ref{eq:montecarloeq} calculates the second-order correlation
function under the assumption of only one atom in the cavity at any time; it
draws upon the one-atom $g^{(2)}(\tau)$ only. There are other cases where an
$N$-atom correlation function can be written as a sum of one-atom correlation
functions. Specifically, when $N$ independent atoms emit simultaneously, the
second-order correlation function of their total emitted field is a sum of
terms involving the one-atom $g^{(1)}(\tau)$ as well as the $g^{(2)}(\tau)$
\cite{carmichael78,hennrich05}. Since we assume there is never more than one
atom in the cavity, and the atoms are separated by $t\gg 1/\kappa$, terms
involving $g^{(1)}(\tau)$ do not appear in Eq.~\ref{eq:montecarloeq}. We note
that Eq. (11) of Ref.\cite{carmichael78} yields our result when $\bar N\ll 1$.


 Figure~\ref{fig:montecarlo} shows an example of the autocorrelation function obtained from
    Eq.~\ref{eq:montecarloeq}.  The
    beam geometry (1.5 mm hole in the MOT, 4.5 cm distance to the cavity and 2
    mm mirror spacing) give parallel and transverse (to the cavity axis)
    divergence angles of $\Delta\theta_p\approx 1/40$ and
    $\Delta\theta_t\approx 1/80$ respectively. The distribution of speeds is
    assumed to be Gaussian with mean $\overline{v}=15$m/s and width $\Delta
    v=1.1$m/s. The angle distribution is assumed to be triangular, as is the
    case for a low-velocity source of atoms from magneto-optical traps
    (MOT)~\cite{lu96}.


If the variation of the cavity-atom coupling is small between adjacent maxima
and minima ({\it i.e.}, $1/t_j\ll 2 \delta$), then
Eq.~(\ref{eq:predictability}) still holds. For each atomic trajectory, under weak drive and with only one atom in the cavity at a time, one may calculate the visibility over the first two
oscillations of $g^{(2)}(\tau)$ before the transit time affects the visibility significantly.
Under these conditions, the quantum beats are in phase, regardless of the path and speed of the atom. Averaging trajectories with the same visibility does not change the final result.

The visibility of the quantum beats is maximal and
constant for a weak drive. Increasing the drive strength brings a decrease in the visibility
and an increased decay rate of the oscillations. Given a photon detection at
$\tau=0$, the probability for detecting the very next photon still exhibits
quantum beats; however, uncorrelated photons contribute at longer
delays. These, for example, might come from the repeated absorption and
emission of $V$- and $H$-polarized photons, respectively; such events
reinitialize the quantum beat with altered phase; higher drive strength
increases their number and decreases the visibility. Also, spontaneous
emission into modes other than those of the cavity may cause decay from
$|e_0\rangle$ to the ground state without providing a heralding photon in the
undriven cavity mode. Following such an emission, the atom might be reexcited
and returned to $|b_0\rangle$, now emitting a photon into the undriven cavity
mode. This photon, not being preceded by a correlated ``start'', further
diminishes the visibility. When the driving is sufficiently weak, the number
of uncorrelated detections approaches zero and maximum visibility is obtained.

To ensure that the decay of the quantum beats is due to the transit of the
atom across the cavity rather than a strong drive, our model shows, for the
parameters used, that the number of photons inside the cavity must be less
than approximately four times the saturation photon number.

We have also verified that so long as any detuning between the $\pi$ atomic
transition (or the drive) and the cavity frequency is smaller than the cavity
width, the frequency of the quantum beats is unaltered. Furthermore, for a
weak drive, unequal Zeeman shifts in the ground and excited states
($\delta\neq\delta^\prime$) do not significantly alter the quantum beat
frequency; it reflects the Larmor frequency of the ground state.

Of more concern from an experimental point of view is the possible presence of
birefringence in the cavity. This works as a direct coupling between its two
orthogonally polarized modes. We model it by introducing such a coupling term
in the system Hamiltonian [Eq.~(\ref{eq:hamiltoniano})].  The most important
differences that this modeling shows are the disappearance of the antibunching
around $\tau=0$ and a change of the quantum beat frequency from twice the
Larmor frequency to the Larmor frequency itself.  The frequency change occurs
when the field generated by birefringence is approximately half the undriven
field generated by the atom.

A final important issue is the effect of two or more atoms in the cavity at a
time. A thorough treatment is reserved for future work. We have, however,
obtained some preliminary indications. Restricting the atomic basis to $N$-
atom symmetric states, we find, in the limit of weak driving and for the
parameters used throughout this paper, that neither the frequency nor duration
of the quantum beats depends on $N$. This follows because the underlying beat
mechanism is independent of $N$.

\section{Conclusions}
We have shown in this paper that a cavity QED system comprising two
orthogonally polarized modes coupled to an atom with magnetic structure shows
robust quantum beats at twice the Larmor frequency of the atomic ground
state. The beats are only visible through conditional measurement of the
intensity of the undriven mode, {\it i.e.}, as photon correlations. They allow
for control and feedback of ground state superpositions created on the
detection of a spontaneously emitted ``start'' photon. Realistic experiments
have been proposed which can be implemented in current realizations of cavity
QED in the optical regime. They open a path to new quantum error correction
protocols and also a new class of quantum feedback and control.

Work supported by CONACYT, M{\'e}xico, NSF of the United States, and the
Marsden Fund of RSNZ of New Zealand.

\end{document}